\documentclass[aps,showpacs,amsmath,amssymb,onecolumn,superscriptaddress,notitlepage,longbibliography,preprintnumbers,prx]{revtex4-2}

\usepackage[english]{babel}
\usepackage{comment}

\usepackage[T1]{fontenc}
\usepackage[utf8]{inputenc}
\usepackage{lmodern}
\usepackage{amsmath,amssymb,amsthm,mathtools,bm,mathrsfs}
\usepackage{bbm}
\usepackage{microtype}
\usepackage{xcolor}
\usepackage{enumitem}
\usepackage{graphicx}
\usepackage[colorlinks=true, allcolors=blue]{hyperref}
\usepackage[table, dvipsnames]{xcolor} 
\usepackage{comment}

\usepackage{subcaption}
\usepackage{cleveref}
\usepackage{dsfont}
\usepackage{multirow}
\usepackage{booktabs}
\usepackage{orcidlink}
\usepackage[normalem]{ulem}
\usepackage{ragged2e}

\newcommand{\ket}[1]{\left|#1\right\rangle}
\newcommand{\bra}[1]{\left\langle#1\right|}
\newcommand{\braket}[2]{\left\langle#1|#2\right\rangle}

\begin{document}

\title{Efficient quantum state preparation on Quantinuum hardware}
\author{Archie Butterworth}
\affiliation{Centre for Quantum Information, 	
Simulation and Algorithms, \\ 
The University of Western Australia, Perth, Australia}
\author{Josh Green}
\affiliation{Centre for Quantum Information, 	
Simulation and Algorithms, \\ 
The University of Western Australia, Perth, Australia}
\author{Yusen Wu}
\affiliation{Centre for Quantum Information, 	
Simulation and Algorithms, \\ 
The University of Western Australia, Perth, Australia}
\author{Jie Pan}
\affiliation{Centre for Quantum Information, 	
Simulation and Algorithms, \\ 
The University of Western Australia, Perth, Australia}
\author{Jingbo Wang}
\email{jingbo.wang@uwa.edu.au}
\affiliation{Centre for Quantum Information, 	
Simulation and Algorithms, \\ 
The University of Western Australia, Perth, Australia}


\begin{abstract}
Preparation and verification of specific quantum states is an important capability for quantum devices to realise advantages over classical computations and algorithms. In this work, we have demonstrated an end-to-end framework that combines resource-efficient quantum state preparation with rapid, robust fidelity verification on near-term quantum hardware. By experimentally preparing and validating a  structured complex quantum state encoding a digitized acoustic signal on the Quantinuum H2-1 trapped-ion platform, we achieved a high hardware fidelity of $F_{\mathrm{hw}} = 0.929$. Crucially, this milestone was realized without relying on idealized assumptions or deep fault-tolerant overhead, but rather through resource-minimal circuits optimized for NISQ-era and early fault-tolerant devices.
Furthermore, we addressed a key limitation in current quantum state certification. While validation methods like shadow overlap work well for random states, their sample complexity can become prohibitively high for the structured states used in practical algorithms. We mitigate this by introducing a pre-measurement basis-change technique that reduces the verification parameter, $\tau$, by over 10 orders of magnitude for structured targets. This approach tightens the theoretical certification guarantees of the shadow overlap method and integrates tensor-network preparation and shadow validation into a unified workflow. These results shift the paradigm of how structured classical data can be mapped to and verified on quantum hardware under realistic noise and measurement budgets. By compressing a robust verification procedure to just 1,000 measurement shots, this framework offers an immediate, scalable benchmarking standard. 

\end{abstract}

\maketitle

\section{Introduction}

Accurate preparation of quantum states is a fundamental prerequisite for the successful implementation of quantum image processing~\cite{venegasandraca2003storing,le2011flexible,zhang2013neqr}, quantum machine learning~\cite{biamonte2017quantum,rebentrost2014quantum,schuld2015introduction,havlicek2019supervised}, linear algebra~\cite{harrow2009quantum,childs2017quantum,gilyen2019quantum}, and quantum chemistry algorithms~\cite{aspuruguzik2005simulated,peruzzo2014variational,mcclean2016theory,mcardle2020quantum,bauer2020quantum,fomichev2024initial}. The primary objective of quantum state preparation is to maximise the accuracy of the prepared state, measured by its fidelity with the target, while minimising total quantum resources. Equally challenging is the successful validation of the fidelity of the state once prepared on an imperfect quantum computer. Since accurate quantum state preparation is imperative for downstream tasks, these challenges represent bottlenecks for the experimental realisation of quantum algorithms. In this work, we propose an end-to-end state preparation and validation scheme aimed at minimising total quantum resource overhead. The proposed approach is experimentally realised on Quantinuum's H2-1 trapped-ion quantum computer.

Recent advances have positioned tensor-network based strategies as leading candidates for efficient and scalable state preparation with minimal quantum resource overhead~\cite{Szoldra2026-ki,Schon2005,Rudolph2022,BenDov2024,Iaconis2024,Malz2024,Smith2024,Melnikov2023}. The limited quantum overhead and shallow circuits make these approaches very well-suited to near-term implementation, including on existing quantum hardware. In addition, tensor networks have recently been explored for the efficient representation of structured classical data~\cite{Lu2021-zr,Jobst2023-rq}. Further, such target states can be efficiently computed using tensor cross interpolation techniques, even for large numbers of qubits \cite{Oseledets2010-xk}. For experimental realisation, we employ a methodology based on the classically optimised disentangling and (by reversal) preparation of matrix product states via shallow quantum circuits consisting of 2-qubit gates \cite{Green2025-pe}. The employed methodology is explicitly designed to minimise circuit depth without the need for ancillas, and is therefore highly promising for the near-term preparation of classical data and other target quantum states.

One of the primary challenges with the experimental realisation of quantum state preparation is to efficiently validate the successful preparation of the encoded target state $\ket{\psi}$.  An ideal certification procedure will use only a small number of local measurements on the prepared state $\rho$ in order to certify its fidelity with the target state $\ket{\psi}$. This is a fundamental yet challenging task, as demanding only local measurements on qubits makes probing the global entanglement structure of the prepared state difficult. Indeed, past methods have either required deep quantum circuits \cite{Huang2020-si,ODonnell2016-at,O-Donnell2017-ey,Haah2016-zq}, exponentially many single-qubit gates \cite{Flammia2011-ac,da-Silva2011-uu,Aolita2015-xz}, or apply only to restricted families of target states \cite{Gluza2018-jx,Takeuchi2018-bb}. The task of validating the fidelity of a prepared state is especially pertinent to current and near-term quantum computation, where limited circuit depth, measurement budgets, and hardware noise make full state tomography infeasible, yet reliable state preparation remains a prerequisite for downstream tasks. These challenges make many proposed fidelity validation techniques impractical for near term hardware. However, the method known as ``shadow overlap'' proposed by \textcite{Huang2025-bh} was developed to address these challenges. The effectiveness of the shadow overlap method is governed by a target-state-dependent parameter $\tau$, which for the structured states most relevant in practice becomes prohibitively large.

Whilst \cite{Huang2025-bh} show that $\tau$ scales efficiently for ``almost all'' states, the proposed method may not cover scenarios in the worst-case. In fact, a Haar random quantum state possesses fundamentally different statistical properties compared with quantum states with specific structures, such as ground states satisfying area laws of entanglement entropy. Direct numerical computation of this parameter reveals that the highly structured target states we wish to prepare exhibit pathologically high $\tau$ values, which in theory render the shadow overlap procedure ineffective for certifying such states. In this work we overcome this limitation by introducing a change-of-basis stage prior to the shadow overlap procedure, reducing $\tau$ by more than $10$ orders of magnitude. Furthermore, the optimal basis can be computed efficiently for matrix product states, making the approach naturally compatible with the tensor-network preparation method used in this work. To empirically demonstrate the end-to-end tensor-network-based approach, we prepare a segment of an acoustic signal on 13 qubits via a circuit consisting of 229 single- and 108 two-qubit gates. We certify the success of the state preparation using the shadow overlap method, reducing uncertainty in the measured fidelity via the proposed basis-change extension to the technique.

\section{Preliminaries and Related Work}
\subsection{Matrix Product States}
Consider a one-dimensional quantum system of $n$ sites with local Hilbert space $\mathbb{C}^2$ and computational basis $\{ \ket{0},\ket{1} \}$. The total Hilbert space dimension is therefore $N=2^n$. A matrix product state (MPS) with open boundary conditions and virtual dimensions $\{\alpha_j\}_{j=0}^n$ (with $\alpha_0 = \alpha_n = 1$) is defined as the variational ansatz \cite{orus2014practical,schollwock2011dmrg}
\begin{equation}
\ket{\psi}
=
\sum_{i_1,\dots,i_n=0}^{1}
\sum_{\{\alpha_j\}}
A^{[1] i_1}_{\alpha_0,\alpha_1}
A^{[2] i_2}_{\alpha_1,\alpha_2}
\cdots
A^{[n] i_n}_{\alpha_{n-1},\alpha_n}
\ket{i_1 i_2 \cdots i_n},
\end{equation}
where for each site $j$ and physical index $i_j$, the tensor
\begin{equation}
A^{[j] i_j}_{\alpha_{j-1},\alpha_j} \in \mathbb{C}^{\alpha_{j-1} \times \alpha_j}
\end{equation}
is a complex matrix.
The bond dimension is defined as
\begin{equation}
\chi := \max_{0\leq j\leq n} \alpha_j,
\end{equation}
which controls the expressive power of the MPS. The total number of variational parameters scales as $O(n \chi^2)$. The MPS structure is central to the state preparation method as well as the efficiency in computing the pre-measurement $\tau$ reduction layer.
\subsection{Schmidt Spectra and Truncation Error}
Given a bipartition of an $n$-qubit system into subsystems $A$ and $B$ (for example,
$A = \{1,\dots,m\}$ and $B = \{m+1,\dots,n\}$), any pure state
$\ket{\psi} \in \mathcal{H}_A \otimes \mathcal{H}_B$ admits a Schmidt decomposition
\begin{equation}
\ket{\psi}
=
\sum_{k=1}^{r} \lambda_k \, \ket{\Phi_k}_A \otimes \ket{\Phi_k}_B,
\end{equation}
where the Schmidt coefficients $\lambda_k \ge 0$ are ordered non-increasingly and satisfy
$\sum_k \lambda_k^2 = 1$. The Schmidt rank
\begin{equation}
r = \mathrm{rank}(\rho_A) = \mathrm{rank}(\rho_B)
\end{equation}
is the rank of the reduced density matrices
$\rho_A = \mathrm{Tr}_B(\ket{\psi}\!\bra{\psi})$ and
$\rho_B = \mathrm{Tr}_A(\ket{\psi}\!\bra{\psi})$ \cite{orus2014practical}.

Consider truncating the virtual bond associated with the bipartition $A|B$ from bond
dimension $\alpha_m$ to $\alpha_m' < \alpha_m$. The Frobenius-norm optimal approximation
is obtained by retaining only the $\alpha_m'$ largest Schmidt coefficients
\cite{Eckart1936-wy, Schmidt1907-vp}. Ordering the coefficients as
$\lambda_1 \ge \lambda_2 \ge \cdots \ge \lambda_{\alpha_m}$, the truncation error is defined as
\begin{equation}
\epsilon_{\mathrm{disc}}(\alpha_m')
:=
\sum_{k > \alpha_m'} \lambda_k^2,
\end{equation}
which is equal to the squared norm of the discarded component of the state.
The fidelity between the original state $\ket{\psi}$ and the (renormalised)
truncated state $\ket{\psi_{\chi'}}$ is then
\begin{equation}
F(\alpha_m') = 1 - \epsilon_{\mathrm{disc}}(\alpha_m').
\end{equation}
\subsection{Schmidt Spectrum Optimisation (SSO)}

The Schmidt Spectrum Optimisation (SSO)  algorithm \cite{Green2025-pe}, is a tensor-network-based procedure for constructing shallow quantum circuits that create an approximate matrix product state (MPS) representation of an $n$-qubit pure state $\ket{\psi}$.

Starting from the target MPS $\ket{\psi^{(1)}}$, the SSO algorithm constructs a sequence of $L$ layers of local unitaries $\{{U_k(\theta_k)}\}_{k=1}^L$ that sequentially remove entanglement from the state. Specifically, the objective at layer $k$ is to optimise $\theta_k$ such that the updated state
\begin{equation}
\ket{\psi^{(k+1)}} = U_k(\theta_k)\ket{\psi^{(k)}}
\end{equation}
has Schmidt spectra $\lambda^{(k+1)}$ that minimises the cost function
\begin{equation}
C\!\left(\lambda^{(k+1)}\right)
=
\sum_{i=1}^{n-1}
\left(
1 -
\left( \lambda^{(k+1)}_{i,1} \right)^2
-
\left( \lambda^{(k+1)}_{i,2} \right)^2
\right).
\end{equation}
where $\lambda^{(k+1)}_{i,1}$ and $\lambda^{(k+1)}_{i,2}$ are the largest and second largest coefficients of the Schmidt decomposition between sites $i$ and $i + 1$, respectively . As described above, this amounts to maximising, at
each bond, the weight retained by the leading two Schmidt coefficients, and hence minimising the truncation error of the $\chi = 2$ MPS approximation to $\ket{\psi^{(k+1)}}$. Each layer consists of the (staircase) circuit ansatz $U_k(\theta_k)$ consisting of a single layer of parameterised $SU(4)$ gates:
\begin{equation}
U_k(\theta_k)
=
\prod_{i=1}^{n-1}
\left(
I^{\otimes (i-1)} \otimes
U_{[i,i+1]}(\theta_{k,i})
\otimes I^{\otimes (n-i-1)}
\right).
\label{layereq}
\end{equation}
which is initialised around the identity. After $L$ iterations of the SSO procedure we obtain
\begin{equation}
\ket{\psi^{(L)}} = U_L \cdots U_2 U_1 \ket{\psi^{(1)}},
\end{equation}
where $\ket{\psi^{(L)}}$ is now able to be well approximated by a $\chi=2$ MPS $\ket{\tilde{\psi}^{(L)}}$. To reverse the disentangling process, we first prepare the low-rank MPS
$\ket{\tilde{\psi}^{(L)}}$ via
\begin{equation}
U_{\mathrm{prep}} \ket{0}^{\otimes n} = \ket{\tilde{\psi}^{(L)}} ,
\end{equation}
where $U_{\mathrm{prep}}$ can be computed exactly by embedding the tensors of $\ket{\tilde{\psi}^{(L)}}$ into a single layer of $SU(4)$ gates, corresponding
to the same circuit structure as a single layer of the staircase ansatz in
Eq. (\ref{layereq}). The full state preparation circuit is then given by
\begin{equation}
U_S := U_1^\dagger U_2^\dagger \cdots U_L^\dagger U_{\mathrm{prep}} ,
\end{equation}
such that $U_S \ket{0}^{\otimes n}$ prepares an approximation of the original target state $\ket{\psi^{(1)}}$.

\subsection{Shadow Overlap Certification}
\label{shadow overlap}
\textcite{Huang2025-bh} propose a method of certifying the fidelity of an unknown $n$-qubit state $\rho$ with a target state $\ket{\psi}$. Using only single-qubit measurements on $\rho$, a quantity known as the shadow overlap $\omega$ is computed for each measurement taken on the circuit. This is done by first uniformly selecting a random $k \in \{0,..,n-1\}$. All qubits are then measured in the $Z$-basis (computational basis) apart from the $k$'th qubit which is measured with equal probability in either the $X,$ $Y,$ or $Z$ pauli basis. The measurement results of all but the $k$'th qubit are stored as a bitstring $z \in \{0,1\}^{n-1}$, and the measured state of the $k$'th qubit in its respective basis is stored as $\ket{s}$. We then classically reconstruct a model of the target state's $k$'th qubit, conditional on the measurement $z$, as
\begin{equation}
\ket{\Psi_{k,z}}:= \frac{1}{\sqrt{|\Psi(z^{(0)})|^2+|\Psi(z^{(1)})|^2}}\left[\Psi(z^{(0)})\ket{0}+\Psi(z^{(1)})\ket{1}\right]
\end{equation}
where $z^{(a)}$ is the binary string that equals $a \in \{0,1\}$ on the $k$'th qubit and $z \in \{0,1\}^{(n-1)}$ in the remaining $n-1$ qubits. The shadow overlap is then computed using the observed state $\ket{s}$ of the $k$'th qubit as 
\begin{equation}
\omega := \langle \Psi_{k,z} \rvert \, (3 \lvert s\rangle\langle s\rvert - \mathbb{I}) \, \lvert \Psi_{k,z} \rangle.
\end{equation}
In practice, $M$ multiple measurements of a circuit are made to compute many shadow overlaps generating an estimate $\bar{\omega}=\frac{1}{M}\sum_{i}\omega_i$ of the true value $ \mathbb{E}[\omega] \approx\bar{\omega} $. This shadow overlap, $\mathbb{E}[\omega]$, acts as a surrogate for the fidelity $\bra{\psi}\rho\ket{\psi}$, where the shadow overlap and fidelity are related through
\begin{align}
\label{linking equations}
    \mathbb{E}[\omega] \geq 1-\epsilon &\implies \bra{\psi} \rho \ket{\psi} \geq 1- \tau\epsilon,\\
 \bra{\psi} \rho \ket{\psi} \geq 1- \epsilon &\implies \mathbb{E}[\omega] \geq 1-\epsilon.
\end{align}
 The parameter $\tau$ determines how well $\mathbb{E}[\omega]$ acts as a surrogate for fidelity, and is defined from the target state $\ket{\psi}$ as follows. Given a pure state $\ket{\psi}$, define the classical Markov transition matrix $P$ associated to $\ket{\psi}$ as 
\begin{equation}
\label{pdefine}
P(x,y) =
\begin{cases}
\displaystyle \frac{1}{N}\,\frac{\pi(y)}{\pi(x)+\pi(y)} 
& (x,y)\in E, \\[1.2ex]
\displaystyle \frac{1}{N}\sum_{x' : (x',x)\in E}
\frac{\pi(x)}{\pi(x)+\pi(x')}
& x = y, \\[1.2ex]
0 & \text{otherwise.}
\end{cases}
\end{equation}
We can then define
\begin{equation}
\tau := \frac{1}{1 - \lambda_2(P)},
\end{equation}
where $\lambda_2(P)$ is the second-largest eigenvalue of $P$. $\tau$ is interpreted as the relaxation time of a random walk on the Boolean hypercube, the graph whose vertices are the set $\{0,1\}^n$, and whose edges are all bit strings differing on exactly one bit. This random walk is constructed through $P$ to have a stationary distribution $\pi(x):=|\braket{x}{\psi}|^2$ , determined by the target state $\ket{\psi}$.

When using the shadow overlap method, it is desirable to have $\tau$ as small as possible, and indeed much of the work of \cite{Huang2025-bh} goes towards putting an upper bound on $\tau$ for almost all target states. In particular, $\tau$ determines how well the shadow overlap approach can distinguish between the target state and other orthogonal states. It is shown in \cite{Huang2025-bh} that $\mathbb{E}[\omega] = \operatorname{Tr}(L\rho)$, and $L\ket{\psi}=\ket{\psi}$ has the largest eigenvalue, where the operator $L$ shares the same eigenvalues as the transition matrix $P$ defined in Equation~\ref{pdefine}. This means that the spectral gap $1/\tau$ bounds the contribution of components orthogonal to $\ket{\psi}$. Consequently, decreasing $\tau$ suppresses the weight that non-target components of the prepared state $\rho$ can contribute to $\operatorname{Tr}(L\rho)=\mathbb{E}[\omega]$, thereby increasing the validity of the shadow overlap as a witness to $\bra{\psi}\rho\ket{\psi}$.

\section{Methodology}
\subsection{Target State: Acoustic Signal Encoding}

The target state prepared in this work is a segment of an acoustic signal recorded in an anechoic chamber, shown in Fig. \ref{fig:target vs flattned}. A loudspeaker emits a Gaussian-modulated tone at 1024 Hz, with the resulting sound pressure signals recorded at four different microphone locations. The signals are sampled at 48,000 Hz, and are used in time-difference-of-arrival (TDOA) based source localisation, meaning the key information content lies in both the temporal waveform structure and the modulation frequency, features which must be accurately preserved through state preparation.

An $N=2^{13} = 8192$ sample segment of the recorded signal from one of the microphone locations is encoded as real-valued amplitudes across a 13-qubit state
\begin{equation}
    |\psi\rangle = \sum_{i=0}^{N-1} c_i |i\rangle,
\end{equation}
where $s_i$ are the raw signal amplitudes and $c_i \propto s_i$ are the normalised signal amplitudes. This signal exhibits significant structure, and provides a representative baseline for the encoding of acoustic signal data relevant to applications such as the beamforming tasks used in medical ultrasound imaging.
\subsection{Circuit Design}
\begin{figure}
    \centering
    \includegraphics[width=0.7\linewidth]{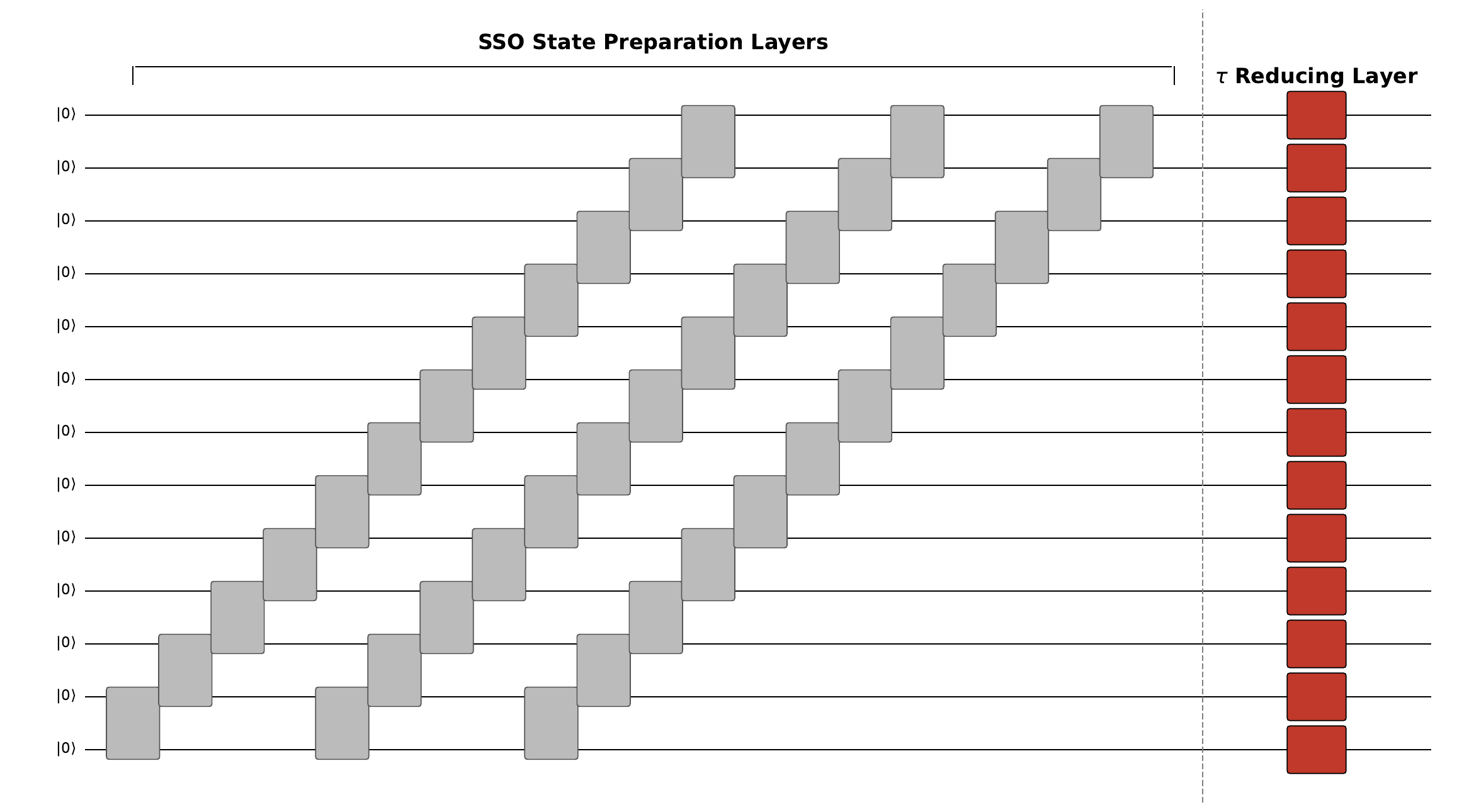}
    \caption{The state preparation circuit consisting of 3 ``staircase'' layers of SU(4) gates, followed by the $\tau$ reducing layer to optimise the shadow overlap certification procedure.}
    \label{fig:circ diagram}
\end{figure}

It is useful to distinguish two sources of infidelity in the end-to-end preparation scheme. The first is the approximation error introduced by the SSO algorithm, which produces an ideal circuit target state $|\tilde{\psi}\rangle = U_S|0\rangle^{\otimes n}$ that approximates but does not exactly reproduce the original signal state $|\psi\rangle$, with classical approximation fidelity $F_{\mathrm{SSO}} = |\langle\psi|\tilde{\psi}\rangle|^2$. 
The second source is hardware noise, which causes the device to produce a mixed state $\rho$ rather than the pure state $|\tilde{\psi}\rangle$. Since $|\tilde{\psi}\rangle$ is our best classical description of the intended circuit output, all certification is performed with respect to this state; the experimentally relevant quantity is therefore $F_{\mathrm{hw}} = \langle\tilde{\psi}|\rho|\tilde{\psi}\rangle$, which isolates hardware noise as the sole source of infidelity in the certification procedure.
Using the Schmidt Spectra Optimisation method \cite{Green2025-pe}, 3 staircase structured SU(4) layers were generated to prepare the target state to a fidelity of $F_{\mathrm{SSO}}=0.9958756$ with the original signal, shown in Fig. \ref{fig:circ diagram}. These SU(4) and SU(2) gates were compiled into the native gate set of Quantiniuum's H2-1 device using Quantiniuum's TKET system into 229 single-qubit ``PhasedX'' gates and 108 2-qubit ``ZZPhase'' gates.

\subsection{The $\tau$ Reducing Layer}

As explained in Section~\ref{shadow overlap}, the correlation between the shadow overlap $\mathbb{E}[\omega]$ and the fidelity $\langle\tilde{\psi}|\rho|\tilde{\psi}\rangle$ is characterised by the value of the parameter $\tau$, the relaxation time of a Markov Chain constructed from the target state $\ket{\tilde{\psi}}$ \cite{Huang2025-bh}. Unlike entanglement and other functions of the Schmidt spectra used in the state preparation, the Markov chain construction, and therefore the value of $\tau$, is basis dependent and can vary by many orders of magnitude for the same quantum state considered in a different basis. This motivated the addition of a layer of single qubit gates designed to change basis in an additional pre-measurement stage, minimising $\tau$ before measurement. The loss function that the single qubit gates were minimised over was the Inverse Partition Ratio, (IPR) a measure of spread for a quantum state \cite{Liu2025-rd}, defined for a normalised state
\begin{equation}
    |\psi\rangle = \sum_{i=0}^{N-1} c_i \, |i\rangle,
\end{equation}
as
\begin{equation}
    \mathrm{IPR}(\psi) = \sum_{i=0}^{N-1}|c_i|^4.
\end{equation}
States with significant structure embedded into the state amplitudes will often have large $\tau$ values, as the resulting probability distribution constructed on the hypercube is typically highly non-uniform and disconnected, leading to congestion and slow mixing times. Minimising the IPR will push the state into a configuration where the amplitudes $|c_i|$ are as close to uniform as possible. This is achievable through single qubit gates, as to do this we do not need to change the entanglement structure of the state. Both the IPR and $\tau$ are independent of the complex phase of each $c_i$ in which the entanglement structure remains, whilst the amplitudes are smoothed out, resulting in a lower $\tau$ value. We show in Appendix~\ref{app:min_ipr_tau} that the minimum value of $\tau$ over all $n$-qubit pure states is $\tau = n$, attained by the uniform superposition, which simultaneously minimises the IPR. This connection motivates the use of IPR minimisation as a surrogate for $\tau$ reduction. Moreover, for an acyclic tensor network target state, the $\tau$ reducing layer can be computed rapidly using classical tensor network optimisation.

\begin{figure}
    \centering
    \includegraphics[width=1\linewidth]{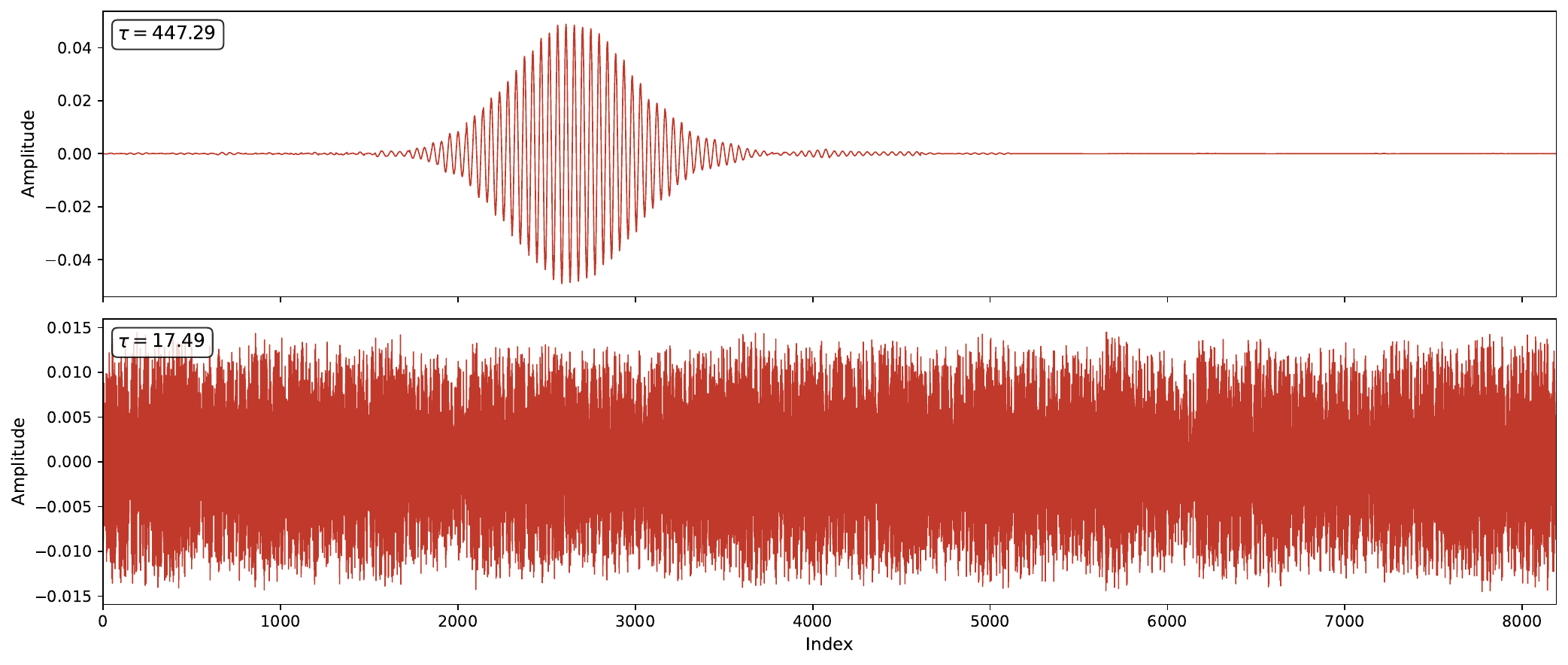}
    \caption{TOP: The digitised acoustic pressure signal encoded in the circuit to produce $\ket{\tilde{\psi}}$. BOTTOM: The same state expressed in the optimised measurement basis after applying the $\tau$ reducing layer, which redistributes the probability amplitudes to reduce $\tau$ from $447.29$ to $17.49$, tightening the theoretical guarantees of the 
shadow overlap certification procedure.}
    \label{fig:target vs flattned}
\end{figure}

Fig. \ref{fig:target vs flattned} shows the effect of the $\tau$ reducing basis change on the acoustic pressure signal target state, where $\tau$ is reduced from $\tau=447.29$ to $\tau=17.49$ through this change of basis. We also explore the effectiveness of this method for more general classes of structured target states in Section~\ref{effective_of_tau}.


\section{Results}
\subsection{Circuit Implementation on Quantinuum hardware}
\label{sec:hardware}
\begin{figure}
    \centering
    \includegraphics[width=0.6\linewidth]{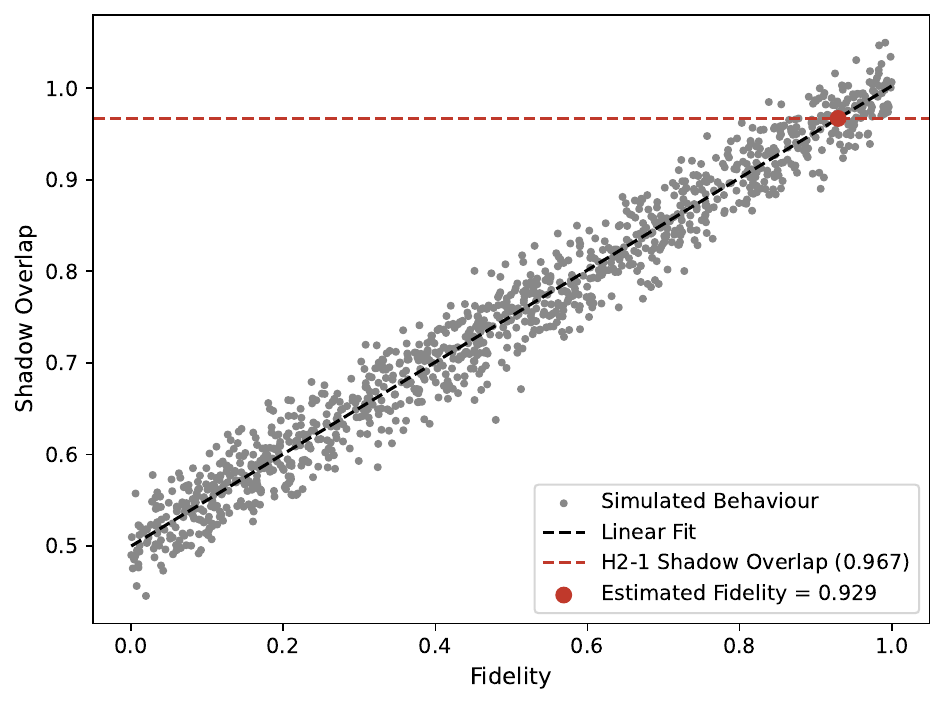}
    \caption{Simulation of the shadow overlap process on randomly prepared states with differing fidelity to the target state. 1000 shots were taken on each prepared state. The experimental value of the shadow overlap obtained from the H2-1 hardware is also plotted, giving an estimate of fidelity as $F_{\mathrm{hw}} = \langle\tilde{\psi}|\rho|\tilde{\psi}\rangle = 0.929$ from a shadow overlap result of $\bar{\omega}_{\text{H2-1}}=0.967$.}
    \label{fig:classical simulation results}
\end{figure}

Before testing on the H2-1 hardware or emulation software, the shadow overlap method with the $\tau$ reduction layer is simulated in the noise-free model for circuits preparing states with varying fidelity to the target state. These ``incorrect'' states were prepared as follows, first a complex random vector $\ket{\eta} \in \mathbb{C}^{2^n}$ is drawn with independent real and imaginary parts sampled from a standard normal distribution. For a desired approximate fidelity $\beta \in [0,1]$, we construct the state
\begin{equation}
\ket{\phi} = \sqrt{\beta}\,\ket{\tilde{\psi}} + \sqrt{1-\beta}\,\ket{\tilde{\eta}},
\end{equation}
where $\ket{\tilde{\eta}}$ is the normalised $\ket{\eta}$. This state $\ket{\phi}$ is prepared in a simulated circuit, and it's fidelity with the target is recorded, along with the estimated shadow overlap $\bar{\omega}$ generated from a simulation of 1000 measurements. The results of this are shown in Fig. \ref{fig:classical simulation results}, which numerically supports what Eqs. (\ref{linking equations}) formalise; that the shadow overlap is indeed a faithful surrogate to fidelity, and that observation of high shadow overlap corresponds to high fidelity.

The state preparation circuit, including the $\tau$ reduction layer shown in Fig. \ref{fig:circ diagram}, was executed on the Quantinuum H2-1 trapped-ion quantum computer, and its corresponding noisy emulator. For each experiment, 1000 shots were taken following the shadow overlap procedure \cite{Huang2025-bh}.

\begin{figure}
    \centering
    \includegraphics[width=0.6\linewidth]{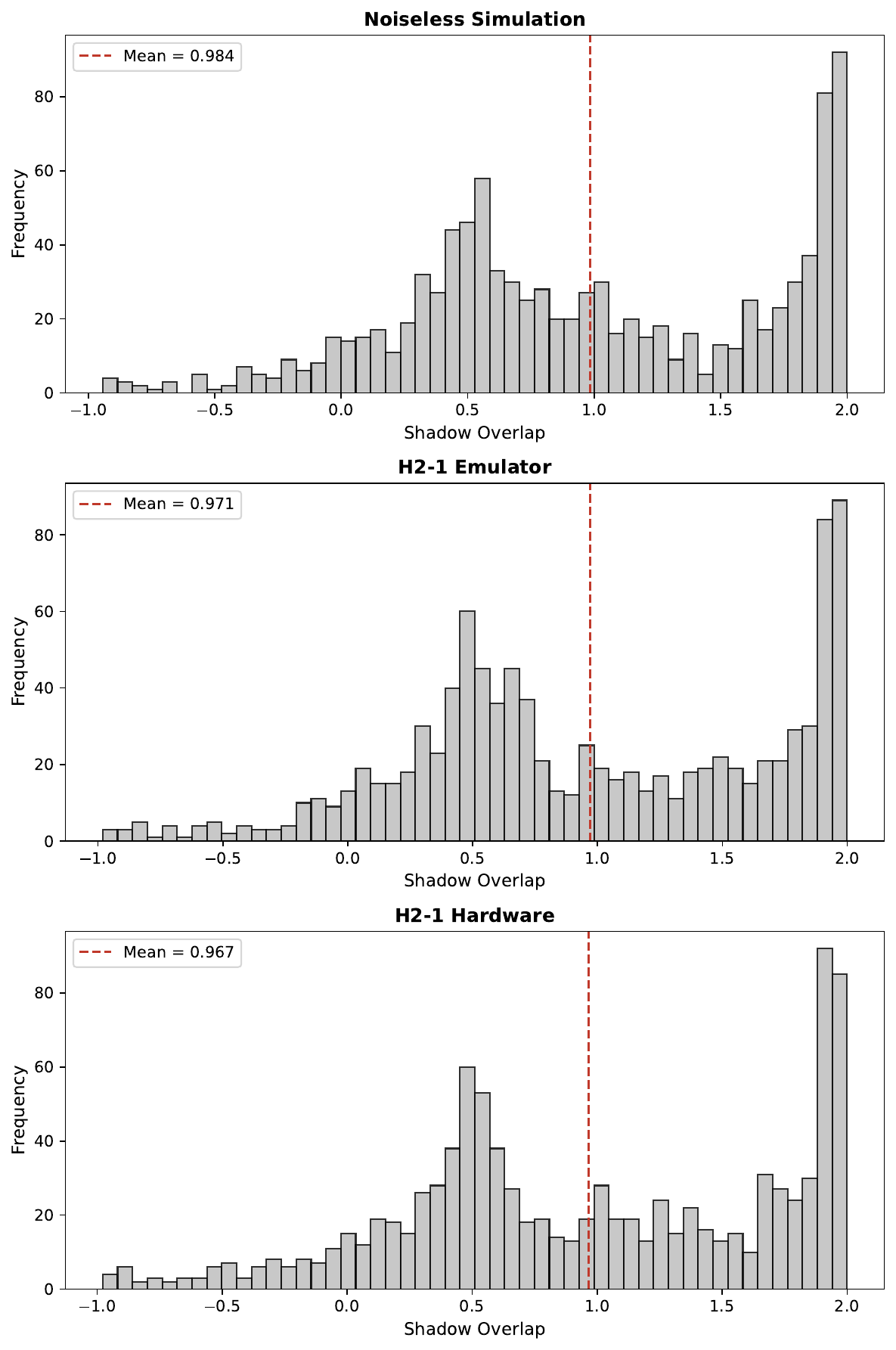}
    \caption{Shadow overlap results from ideal noise-free simulation, noisy emulation and the H2-1 trapped ion device.}
    \label{fig:quantinium reults}
\end{figure}

The shadow overlaps calculated from these measurements on H2-1 are shown in Fig. \ref{fig:quantinium reults}, alongside noisy emulator and noiseless simulation results. In particular, we obtain estimates as $\bar{\omega}_{\text{H2-1}}=0.967$ on the H2-1 trapped ion quantum computer,  $\bar{\omega}_{\text{H2-1E}}=0.971$ on the corresponding emulator and $\bar{\omega}_{\text{Ideal}}=0.984$ under ideal simulation. Using this, alongside the simulation results shown in Fig. \ref{fig:classical simulation results}, we are able to estimate the fidelity of the prepared state to the ideal circuit target as $\langle\tilde{\psi}|\rho|\tilde{\psi}\rangle=0.929$.

The estimated hardware fidelity $F_{\mathrm{hw}} = 
\langle\tilde{\psi}|\rho|\tilde{\psi}\rangle = 0.929$ is obtained by certifying the prepared state $\rho$ against the ideal SSO circuit target $|\tilde{\psi}\rangle$, isolating hardware noise as the sole contributor to this infidelity. Separately, the SSO algorithm achieves a classical approximation fidelity of $F_{\mathrm{SSO}} = 
|\langle\psi|\tilde{\psi}\rangle|^2 = 0.9958756$ between the circuit target and the original signal state $|\psi\rangle$. The gap between $F_{\mathrm{hw}} = 0.929$ and the ideal value of unity is therefore attributable entirely to hardware noise on the H2-1 device, rather than to any limitation of the state preparation algorithm. This interpretation is supported by the close agreement between the noisy emulator result $\bar{\omega}_{\mathrm{H2\text{-}1E}} = 0.971$ and the hardware result $\bar{\omega}_{\mathrm{H2\text{-}1}} = 0.967$, suggesting the emulator faithfully captures the noise behaviour of the device.
\subsection{Effectiveness of $\tau$ reducing layer}
\label{effective_of_tau}

Here a more detailed exploration of the $\tau$ reducing method is given. The effectiveness of this technique depends on two key points, that the IPR acts as a suitable indicator of $\tau$, and that it is possible to reduce the IPR and therefore $\tau$ on the types of states relevant to state preparation. This is explored numerically, finding that the IPR acts as a strong surrogate for $\tau$, and that a change of basis will greatly reduce $\tau$ by many orders of magnitude for states with pathologically high $\tau$ values.
It is important to note that even in the case where one cannot directly compute $\tau$ due to a high dimensional target state, the IPR method can still and should be effectively implemented with the knowledge that it will protect against pathologically high $\tau$ values, allowing for the shadow overlap method to be used. Additionally, as the layer of single qubit gates can be absorbed into the staircase structure of the state preparation algorithm, there is no additional circuit depth induced through this strategy.

We explore the effectiveness of the $\tau$ reducing layer on randomly generated structured target states. Simply generating Haar Random states does not generate a \textit{structured} target state and instead results in low IPR and low $\tau$ states; these states are not the states of interest for the $\tau$ reducing method as they already display mixing times of the same order of magnitude of the minimum possible $\tau$. Indeed, for any $n$-qubit state $\ket{\psi}$, the smallest possible $\tau$ constructed over the full $n$-dimensional hypercube is $\tau=n$. This is achieved by any such $\ket{\psi}$ for which all probability amplitudes are equal, and also corresponds to the minimum IPR value, the proof of which can be found in the appendix. 

\begin{figure}
    \centering
    \includegraphics[width=0.7\linewidth]{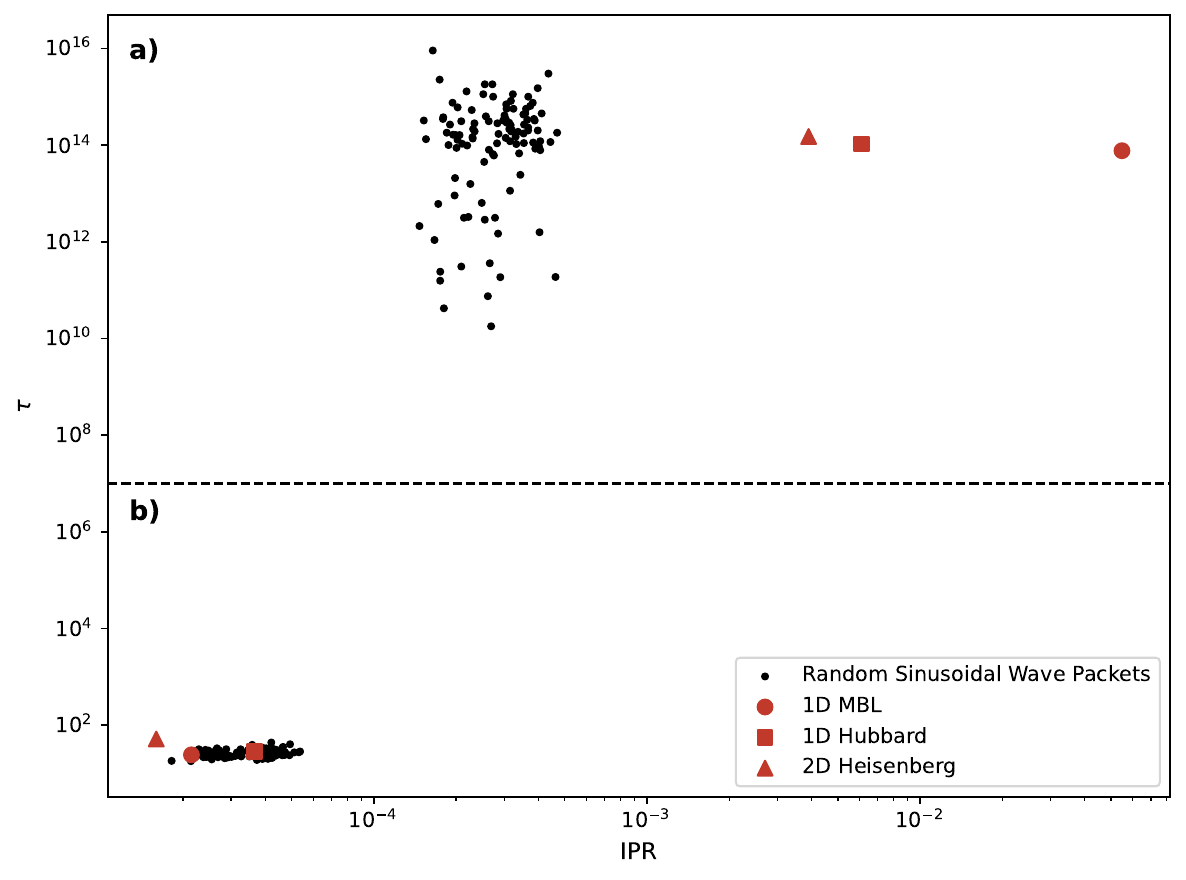}
    \caption{The values of $\tau$ and IPR plotted before a), and after b), the action of the $\tau$ reducing layer, for randomly generated wave-packets and 3 ground states of lattice hamiltonians. In all cases the $\tau$ reducing layer is able to bring down $\tau$ by at least 10 orders of magnitude.}
    \label{fig:beforeandaftertau}
\end{figure}

In order to explore the behaviour of some of the structured states that desired state preparation signals are expected to resemble, we generated sinusoidal wave packets in $n=16$ qubit states with high $\tau$ values to test the effectiveness of the $\tau$ reducing layer on. Fig. \ref{fig:beforeandaftertau} shows the effectiveness of this process, reducing $\tau$ for these states by at least 10 orders of magnitude. The $\tau$ reducing layer is able to effectively ``scramble'' the structured states into a new basis in which $\tau$ remains stable at the order of magnitude of $\mathcal{O}(n)$. We see that forcing the IPR to be lower in turn reduces $\tau$, supporting the choice of using the IPR as an efficiently computable surrogate for the more computationally intensive parameter $\tau$.

Alongside the randomly generated sinusoidal states we also demonstrate the power of the $\tau$ reducing layer on 3 different ground states of lattice hamiltonians, (the same states shown in \cite{Green2025-pe} to be effective under the SSO framework). The states tested were:
\begin{enumerate}
\item \textbf{1D Many-Body Localised (MBL) Spin Chain:}  
The ground state of an $n=14$ site Spin-1/2 chain with random longitudinal fields $h_i \sim \mathcal{N}(0,1)$ and open boundaries.

\item \textbf{1D Spinless Hubbard Chain:}  
The ground state of the spinless Hubbard chain with $n = 14$
sites, hopping amplitude $t = 0.5$, nearest–neighbour interaction strength $V = 1.0$, and chemical potential $\mu = 1.0$, with open boundary conditions.

\item \textbf{4$\times$4 2D Heisenberg Model (Spin-1/2):}  
The ground state of the spin-1/2 2D Heisenberg model on a 4 × 4 square lattice with nearest–neighbour coupling J = 1.0 and open boundary conditions
\end{enumerate}

The effect of the $\tau$ reducing layer on these states are displayed alongside the randomly generated sinusoidal states in Fig \ref{fig:beforeandaftertau}. Like the sinusoidal states, these ground states of lattice hamiltonians also have exceptionally large tau, and benefit greatly from the basis change. This is the primary function of the $\tau$ reducing layer: to protect against pathological states where $\tau$ is many orders of magnitude higher than expected, improving the validity of the shadow overlap process.

\section{Conclusions}

In this work we have demonstrated an end-to-end quantum state preparation 
and certification scheme on the Quantinuum H2-1 trapped-ion quantum 
computer, preparing a 13-qubit state encoding a digitised acoustic pressure 
signal. The SSO algorithm achieves a classical approximation fidelity of 
$F_{\mathrm{SSO}} = 0.9958756$ between the circuit target $\ket{\tilde{\psi}}$ 
and the original signal state $\ket{\psi}$, and the hardware preparation 
fidelity to this ideal prepared state $\ket{\tilde{\psi}}$is estimated as $F_{\mathrm{hw}} = \langle\tilde{\psi}|\rho|
\tilde{\psi}\rangle = 0.929$. The close agreement between hardware and emulator shadow overlap values, 
($\bar{\omega}_{\mathrm{H2\text{-}1}} = 0.967$ and 
$\bar{\omega}_{\mathrm{H2\text{-}1E}} = 0.971$ respectively), confirms 
that the emulator faithfully captures the dominant noise processes of 
the device.

A central contribution of this work is the $\tau$ reducing layer, 
a pre-measurement basis change that tightens the theoretical 
certification guarantees of the shadow overlap method for structured 
target states. By minimising the IPR as a tractable surrogate for $\tau$, the relaxation 
time of the Markov chain associated with the target state, we reduce $\tau$ 
from $447.29$ to $17.49$ for our acoustic target state, and demonstrate 
reductions of at least 10 orders of magnitude for a broad class of 
structured states including sinusoidal wave packets and ground states of 
lattice Hamiltonians, tightening the theoretical 
sample complexity bounds of the shadow overlap method and providing 
robustness against pathological failure cases that would otherwise 
render certification ineffective under the framework proposed by \cite{Huang2025-bh}. Importantly, the $\tau$ reducing layer introduces no additional 
circuit depth, as the single-qubit basis change gates can be absorbed into 
the final layer of the SSO staircase structure upon circuit compilation.

A key practical advantage of the IPR-based approach is that it remains 
viable even when the target state is too large for $\tau$ to be computed 
directly. Since the IPR can be computed efficiently whenever the target 
state admits an MPS representation, the $\tau$ reducing layer provides 
protection against pathologically large $\tau$ values for the structured 
states most relevant to quantum state preparation applications, including 
medical imaging data of the kind considered here.

Future work could explore applying this end-to-end scheme to larger and 
more complex medical datasets, taking advantage of improvements in 
near-term quantum hardware to push towards practically relevant system 
sizes. From a theoretical perspective, the numerical evidence presented 
here suggests that the $\tau$ reducing layer is able to bring $\tau$ 
down to $\mathcal{O}(n)$ for a broad range of structured states. If this 
could be established analytically, it would imply that the shadow overlap 
method is an efficient certification procedure for arbitrary target states, 
resolving a central open question of \cite{Huang2025-bh} and significantly 
broadening the practical applicability of shadow overlap certification.

\section*{Acknowledgement}

This project was supported by the Australian Government through the Critical Technologies Challenge Program (CTCP). Computational resources were provided by the Pawsey Supercomputing Research Centre. We thank Quantinuum for quantum hardware access, with particular thanks to Maud Einhorn, Vincent Anandraj, Joshua Savory, and Sam White for their assistance and technical support. We also acknowledge Sam Marsh (Q-CTRL) and Josh Snow (UWA) for valuable discussions.


\appendix
\section{The minimum IPR corresponds to the minimum $\tau$.}
\label{app:min_ipr_tau}
We wish to show that among all quantum states $\ket{\psi}$ on the $n$-dimensional  boolean hypercube, the uniform superposition minimizes the relaxation time $\tau$ of the associated Markov chain, and that this minimum relaxation time is $\tau = n$. 
Equivalently, since the relaxation time $\tau = 1/\gamma$ is minimized when the spectral gap $\gamma(P)$ is maximized, we seek to show that the transition matrix $P$ corresponding to the uniform distribution $\pi(x) = 1/N$ maximizes $\gamma(P)$ over all $P \in \mathcal{P}$.

We also note that the uniform distribution minimizes the Inverse Participation Ratio (IPR), defined as
\[
\text{IPR}(\psi) = \sum_{x \in \{0,1\}^n} |\psi_x|^4 = \sum_x \pi(x)^2.
\]
By the Cauchy-Schwarz inequality,
\[
\sum_x \pi(x)^2 \geq \frac{1}{N},
\]
with equality if and only if $\pi(x) = 1/N$ for all $x$, i.e. the uniform distribution. Thus the uniform superposition simultaneously minimizes the IPR and, as we will show, minimizes the relaxation time.
Given a quantum state
\[
\ket{\psi} = \sum_{x \in \{0,1\}^n} \psi_x \ket{x},
\]
we associate with it the probability distribution
\[
\pi(x) = |\psi_x|^2.
\]
First we will show that the function
\[
\lambda_2: \mathcal{P}\to\mathbb{R}
\]
that returns the second largest eigenvalue of $P \in \mathcal{P}$ is convex, where $\mathcal{P}$ is the set of all possible transition matrices on the boolean hypercube defined from the state vector as 
\[
P(x,y) =
\begin{cases}
\displaystyle \frac{1}{N}\,\frac{\pi(y)}{\pi(x)+\pi(y)} 
& (x,y)\in E, \\[1.2ex]
\displaystyle \frac{1}{N}\sum_{x' : (x',x)\in E}
\frac{\pi(x)}{\pi(x)+\pi(x')}
& x = y, \\[1.2ex]
0 & \text{otherwise.}
\end{cases}
\]
This is the definition of $P$ that \cite{Huang2025-bh} provides. For any $P \in \mathcal{P}$, let the eigenvalues be $1 = \lambda_1(P) \geq \lambda_2(P) \geq \dots \geq \lambda_n(P)$. The spectral gap is defined as:
\[ \gamma(P) = 1 - \lambda_2(P) \]
By the Courant-Fischer Theorem, for a symmetric matrix $P$, the second largest eigenvalue $\lambda_2(P)$ can be expressed as:

\[ \lambda_2(P) = \sup_{\substack{\|v\|_2 = 1 \\ v \perp \mathbf{1}}} v^T P v \]
where $\mathbf{1}$ is the eigenvector corresponding to $\lambda_1 = 1$. Now, for any $P_1, P_2 \in \mathcal{P}$ and $\alpha \in [0, 1]$:
\begin{align*}
\lambda_2(\alpha P_1 + (1-\alpha) P_2) &= \sup_{v \perp \mathbf{1}} v^T (\alpha P_1 + (1-\alpha) P_2) v \\
&\leq \alpha \sup_{v \perp \mathbf{1}} v^T P_1 v + (1-\alpha) \sup_{v \perp \mathbf{1}} v^T P_2 v \\
&= \alpha \lambda_2(P_1) + (1-\alpha) \lambda_2(P_2)
\end{align*}
This shows  that $\lambda_2(P)$ is convex, so $1-\lambda_2(P)=\gamma(P)$ is concave. Let $\mathcal{G}$ be the automorphism group of the $d$-dimensional hypercube $Q_d$. For any $\sigma \in \mathcal{G}$, let $M_{\sigma}$ be its corresponding permutation matrix. 

Given a transition matrix $P$, define the averaged matrix $\bar{P}$ as:
\[ \bar{P} = \frac{1}{|\mathcal{G}|} \sum_{\sigma \in \mathcal{G}} M_{\sigma} P M_{\sigma}^T \]
Because the action of $\mathcal{G}$ on the hypercube is uniquely determined by the image of one vertex and a permutation of its neighbours, $\bar{P}$ assigns equal weight to every edge. This can be made clear by looking at just one matrix element $P(x,y)$. Any edge $(u,v)$ can be mapped to $(x,y)$ by some $g \in \mathcal{G}$. First fix $g(u)=x$ and $g(v)=y$, then we are free to order the rest of the $n-1$ edges touching $u$ as we want, giving $(n-1)!$ different $g$ that map $(u,v)$ to $(x,y)$. This is true for any 2 edges, so each edge weighting is given by the same summation. Therefore $\bar{P}$ corresponds to the matrix where $\pi(x)$ is the uniform distribution.

As $\gamma(P)$ is a concave function, Jensen's Inequality implies:
\[ \gamma(\bar{P}) = \gamma\left( \frac{1}{|\mathcal{G}|} \sum_{\sigma \in \mathcal{G}} M_{\sigma} P M_{\sigma}^T \right) \geq \frac{1}{|\mathcal{G}|} \sum_{\sigma \in \mathcal{G}} \gamma(M_{\sigma} P M_{\sigma}^T) \]
Since $M_{\sigma} P M_{\sigma}^T$ is a similarity transformation, it preserves the spectrum of $P$. Therefore, $\gamma(M_{\sigma} P M_{\sigma}^T) = \gamma(P)$ for all $\sigma$. The inequality simplifies to:
\[ \gamma(\bar{P}) \geq \frac{1}{|\mathcal{G}|} \sum_{\sigma \in \mathcal{G}} \gamma(P) = \gamma(P) \]

Therefore, transition matrix $\bar{P}$ maximizes the spectral gap $\gamma$ among all possible $P \in \mathcal{P}$ transition matrices on the hypercube. Since the relaxation time is $\tau = 1/\gamma$, this corresponds to the minimum possible $\tau$. 
We can now calculate this minimum $\tau$. When $\pi(x)$ is uniform, \(P\) simplifies to
\begin{equation}
P
=
\frac{1}{2}\, I
+
\frac{1}{2n}\sum_{i=1}^n X_i,
\end{equation}
where \(X_i\) denotes the Pauli-\(X\) operator acting on qubit \(i \in [n]\). The first eigenvalues of \(P\) is $\lambda_0=1$. The second-largest eigenvalue is
\[
\lambda_1
=
1 - \frac{1}{n}.
\]
Therefore, the relaxation time satisfies
\[
\tau
=
\frac{1}{\lambda_0 - \lambda_1}
=
n.
\]

\bibliographystyle{apsrev4-2}
\bibliography{references}
\end{document}